\documentclass[10pt]{iopart}
\usepackage[dvips]{graphicx}
\usepackage{iopams}

\def\cala{{\mathcal A}} 
\def\calc{{\mathcal C}} 
\def\cald{{\mathcal D}}

\newcommand{\bino}[2]{\left(\hspace{-.2cm}\begin{array}{c}#1\\#2\end{array}\hspace{-.2cm}\right)}

\begin{document} 

\title{Purity distribution for bipartite random pure states}
\author{Olivier Giraud} 
\address{Laboratoire de Physique Th\'eorique, 
Universit\'e Toulouse III, CNRS, 31062 Toulouse, France}
\ead{giraud@irsamc.ups-tlse.fr}
\date{\today} 
\begin{abstract}
Analytic expressions for the probability density distribution of 
the linear entropy and the purity are derived 
for bipartite pure random quantum states. The explicit distributions 
for a state 
belonging to a product of Hilbert spaces of dimensions $p$ and $q$
are given for $p=3$ and any $q\geq 3$, as well as for $p=q=4$. 
\end{abstract}
\pacs{03.67.Mn, 03.67.-a}
\maketitle


\section*{Introduction}
Characterizing entanglement is one of the challenging issues that has
been fostered by the development of quantum computation in the past few years. 
Beyond the obvious interest for the foundations of quantum mechanics,
the motivations increased as the role played by entanglement in the power
of quantum algorithms was made clearer.

As random states are entangled with high probability, they play an 
important role in the field of quantum communication, and
appear in many algorithms, such as quantum data hiding protocols 
\cite{DivLeuTer} or superdense coding \cite{HarHayLeu}.
Various quantum algorithms were proposed to generate random states: 
algorithms based on the entangling
power of quantum maps have been proposed in \cite{ZanZalFao}; chaotic
maps were considered for instance in \cite{BanLak, Sco},
pseudo-integrable maps in \cite{GirGeo}. It was also proposed to
generate random states by construction of pseudo-random operators
\cite{WeiHel}, or by sequences of two-qubit gates \cite{Zni07}. 
The efficiency of all these algorithms is measured by their 
ability to reproduce the entangling properties of
random states. Thus, it is crucial to have some measure of 
entanglement.

The problem of quantifying entanglement is a difficult issue, and a number of
entanglement measures have been proposed (see e.g. the review \cite{PleVir}).
However for pure quantum states, bipartite entanglement, that is the entanglement
of a subset of the qubits with the complementary subset, is 
essentially measured by the von Neumann entropy of the reduced density matrix
\cite{BenBerPopSch}. More convenient to use are the linear entropy 
or the purity, which are linearized versions of the von Neumann entropy.
Purity or linear entropy were used to estimate the entangling 
properties of chaotic quantum maps (for instance in the Baker's map
\cite{Sco} or in kicked tops \cite{DemKus}), or 
entanglement growth under random unitary evolution \cite{AbrVal}.
It has been used as a reference to compare the accuracy of various 
entanglement measures \cite{FacFloPas}, or to measure 
dynamical generation of entanglement in coupled bipartite systems  
\cite{ZniPro}.
Recently, it was shown that purity for a pure 
quantum state could be expressed as a function
of the inverse participation ratio, thus enabling to connect entanglement
(as measured by the purity) to localization properties of quantum 
states \cite{GirMarGeo, VioBro}.

In most of these works, the entanglement properties of the systems that
are studied are compared to those of random pure states. 
This is usually done by resorting to numerical
computations, based e.g. on Hurwitz parametrization \cite{Hur} of random
states. Of course, a comparison directly based not on numerics but on 
analytical formulae 
describing the entanglement properties of random pure states 
would be most desirable. In this paper we derive such formulae for the
purity and the linear entropy.

Random pure states can be realized as column vectors of random unitary
matrices drawn from an ensemble with unitarily 
invariant Haar measure \cite{Meh}. Equivalently they can be realized as vectors
with coefficients given by independent random complex Gaussian
variables, rescaled to have a norm equal to 1 (see e.g. \cite{ZycSom}).
Various aspects of the entanglement properties of random pure states 
have been studied in previous works: the distribution of $G$-concurrence has
been calculated in \cite{CapSomZyc}; the average von Neumann entropy
had been obtained in \cite{Pag};
the moments for the purity distribution in random states were calculated
analytically in \cite{Gir07}, as well as approximate moments of Meyer-Wallach
entanglement (a multipartite entanglement measure based on the purity).  
The aim of this paper is to provide analytic expressions for 
the probability density distribution of the purity (or the linear entropy)
for bipartite random pure states. Explicit analytical formulae are
derived for the smallest bipartitions of the Hilbert space into a
$p$-dimensional and a $q$-dimensional spaces: $p=q=3$ in Section
\ref{pder33},  $p=3$ and any $q$ in Section \ref{pder3q},  $p=q=4$ in Section
\ref{pder44}. A different method allowing to obtain formulae for $p=4$ and
 $q\geq 4$ is explained in Section \ref{pder44}. The method is general,
 and similar calculations would yield expressions for higher values 
of $p$ and $q$, although it is not sure whether these formulae would
take any nice and compact form.

\section{Probability density distribution of the purity.}
Let us consider a state $\Psi$ belonging to the Hilbert space 
$\mathbb{C}^p \otimes \mathbb{C}^q$ for some integers $p,q$ with $p\leq q$.
Suppose $\Psi$ admits the following Schmidt decomposition \cite{NieChu}
\begin{equation}
\Psi=\sum_{i=1}^{p}\sqrt{x_i}|a_i\rangle\otimes|b_i\rangle,
\end{equation}
where $|a_i\rangle$ and  $|b_i\rangle$, $1\leq i\leq p$, are respectively 
orthonormal bases for $\mathbb{C}^p$ and 
$\mathbb{C}^q$. The purity $R$ of the state $\Psi$ can be written in
terms of Schmidt coefficients as
\begin{equation}
\label{purity}
R(\Psi)=\sum_{i=1}^{p}x_i^2.
\end{equation}
The linear entropy is expressed in terms of the purity by the simple relation
\begin{equation}
\label{relationRS}
S_L(\Psi)=\frac{p}{p-1}\left(1-R(\Psi)\right).
\end{equation}
For random pure states obtained as column vector of random matrices
distributed according to the unitarily invariant Haar measure (CUE matrices),
the Schmidt coefficients are characterized by the following 
joint distribution \cite{ZycSom}: 
\begin{equation}
\label{density}
P(x_1, \ldots, x_p)=\cala
\prod_{1\leq i<j\leq p}(x_i-x_j)^2
\prod_{1\leq i\leq p}x_i^{q-p}\;\delta\left(1-\sum_{i=1}^{p}x_i\right)
\end{equation}
for $x_i\in[0,1]$; $\delta$ is the Dirac delta function, and $\cala$ is the normalisation factor 
\begin{equation}
\label{normalisation}
\cala=\frac{(p q-1)!}{\prod_{0\leq j\leq p-1}(q-j-1)!(p-j)!}.
\end{equation}
The purity distribution function is then given by
\begin{equation}
\label{defpr}
P(R)=\cala\int_0^1\hspace{-0.3cm} d^p{\bf x} V({\bf x})^2
\prod_{1\leq i\leq p}x_i^{q-p}\delta\left(1-\sum_{i=1}^{p}x_i\right)
\delta\left(R-\sum_{i=1}^{p}x_i^2\right),
\end{equation}
where we have introduced the Vandermonde determinant
$V({\bf x})=\prod_{i<j}(x_i-x_j)$
with ${\bf x}=(x_1, \ldots, x_p)$. 
The distribution $\tilde{P}(S_L)$ of the linear entropy can be straightforwardly 
deduced from $P(R)$ using Eq.~\eref{relationRS}:
\begin{equation}
\tilde{P}(S_L)=\frac{p-1}{p}P\left(1-\frac{p-1}{p}S_L\right).
\end{equation}
As the purity vanishes outside the interval $[1/p, 1]$, the distribution
of linear entropy is supported by $[0,1]$.

In the next sections we 
evaluate \eref{defpr} in the cases $p=2,3,4$. We first note that the
expression  \eref{defpr} can be simplified to 
\begin{equation}
\label{defpr_simplifie}
P(R)=\cala p!
\int_0^1d^p{\bf x} V({\bf x})
\prod_{1\leq i\leq p}x_i^{d+k-1}\delta\left(1-\sum_{i=1}^{p}x_i\right)
\delta\left(R-\sum_{i=1}^{p}x_i^2\right),
\end{equation}
where $d=q-p$, using a transformation detailed in \cite{Gir07}.

\section{Distribution of the purity for a 
$2 \times q$ bipartite random state.}
\label{pder2q}
In the case $p=2, q\geq 2$, the analytic expression for the probability 
distribution $P(R)$ can easily be obtained analytically directly 
by integration of \eref{defpr}. It reads
\begin{equation}
\label{P2}
P(R)=\frac{(2 q-1)!}{2^{q-1}(q-1)!(q-2)!}(1-R)^{q-2}\sqrt{2R-1}
\end{equation}
for $1/2\leq R\leq 1$, and 0 otherwise.

\section{Distribution of the purity for a 
$3 \times 3$ bipartite random state.\label{pder33}}
\begin{figure}[ht]
\begin{center} 
\includegraphics*[width=.85\linewidth]{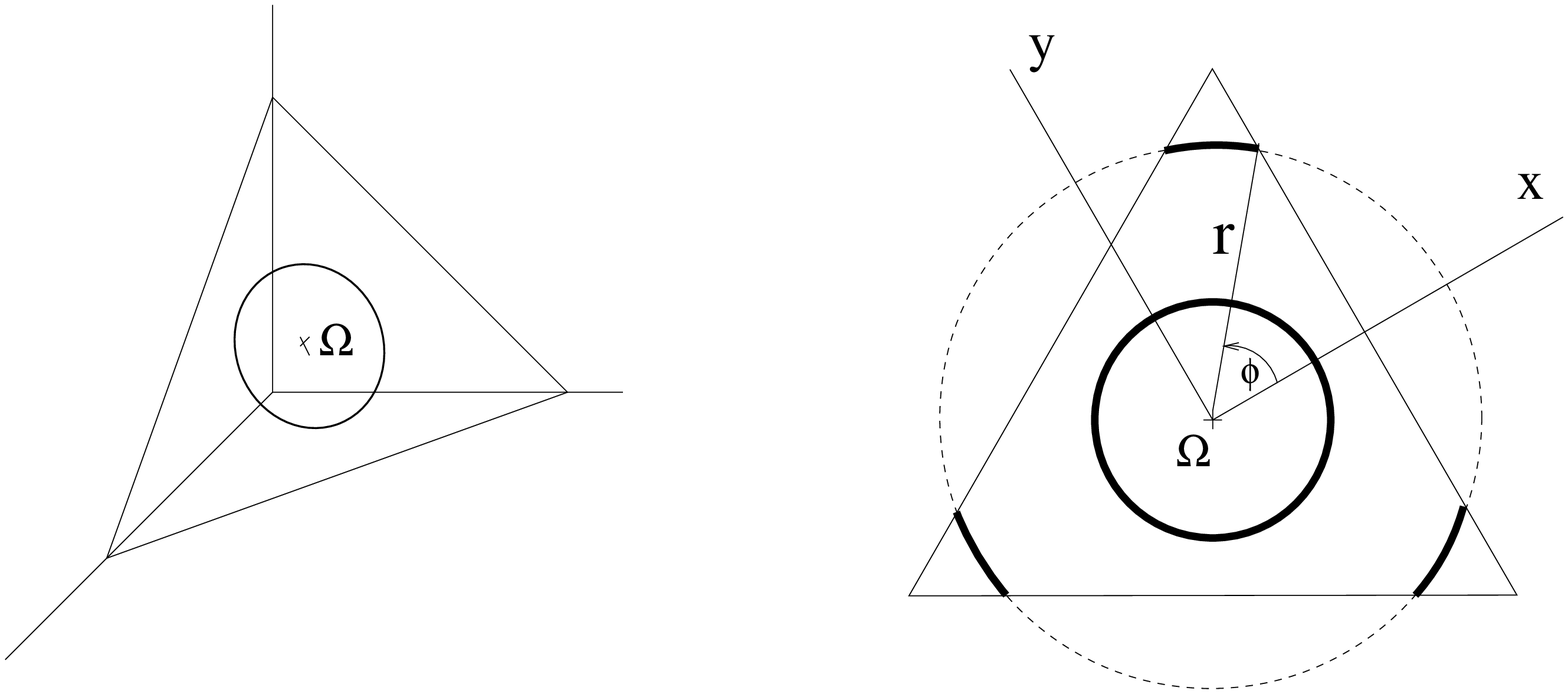}
\end{center} 
\caption{Domain of integration for $p=3$. Representation in $\mathbb{R}^3$ for 
$\frac{1}{3}\leq R\leq \frac{1}{2}$ (left), and in the plane $x_1+x_2+x_3=1$ 
(right, thick lines).}
\label{trianglep3}
\end{figure}

In the case $p=3$ and $q=3$, Eq.~\eref{defpr} reads
\begin{equation}
\label{defpr3}
P(R)=\frac{8!}{4}\int_0^1d^3{\bf x}V({\bf x}) x_2 x_3^2
\delta\left(1-\sum_{i=1}^{3}x_i\right)
\delta\left(R-\sum_{i=1}^{3}x_i^2\right).
\end{equation}
The domain of integration is the intersection of the cube $[0,1]^3$, 
the plane $x_1+x_2+x_3=1$ and the sphere of radius $\sqrt{R}$ 
centered on the origin. 
This intersection is a circle if $\frac{1}{3}\leq R\leq \frac{1}{2}$ 
(see Fig.~\ref{trianglep3} left), or sections of a circle  if $\frac{1}{2}\leq R\leq 1$.
We first make the change of variables ${\bf x}=O{\bf X}$, 
where $O$ is the orthogonal matrix
\begin{equation}
\label{matrixO}
\left(
\begin{array}{ccc}
-\frac{2}{\sqrt{6}}&0&\frac{1}{\sqrt{3}}\\
\frac{1}{\sqrt{6}}&-\frac{1}{\sqrt{2}}&\frac{1}{\sqrt{3}}\\
\frac{1}{\sqrt{6}}&\frac{1}{\sqrt{2}}&\frac{1}{\sqrt{3}}
\end{array}\right).
\end{equation}
In the new coordinates $(X_1, X_2, X_3)$, the plane has equation $X_3=1/\sqrt{3}$ and the sphere
$X_1^2+X_2^2+X_3^2=R$. We first integrate over $X_3$ and make
a change of variables to polar
coordinates $(X_1=\rho\cos\theta,X_2=\rho\sin\theta)$. Let us 
set $\phi=0$ if  $\frac{1}{3}\leq R\leq \frac{1}{2}$, 
$\phi=\arccos(1/\sqrt{6R-2})$ if $\frac{1}{2}\leq R\leq 1$, 
and $r=\sqrt{R-1/3}$. The domain of integration
in the plane  $X_3=1/\sqrt{3}$ is represented in  Fig.~\ref{trianglep3}
right for two different values of $R$. After integrating over $\rho$ we obtain
\begin{eqnarray}
\label{prtheta}
P(R)&=&-\frac{7!r^3}{\sqrt{6}}\int_{\cald_{\phi}} d\theta 
\left(\frac{1}{3}-r\sqrt{\frac{2}{3}}\cos\left(\theta+\frac{4\pi}{3}\right)
\right)\\
\nonumber&\times&
\left(\frac{1}{3}-r\sqrt{\frac{2}{3}}\cos\left(\theta+\frac{2\pi}{3}
\right)\right)^2\sin(3\theta).
\end{eqnarray}
The domain of integration for $\theta$ is 
$\cald_{\phi}=[0,2\pi]\setminus\calc_{\phi}$, where 
\begin{equation}
\calc_{\phi}=\left[-\phi\frac{}{},\phi\right]\bigcup \left[-\phi+\frac{2\pi}{3},
\phi+\frac{2\pi}{3}\right]\bigcup\left [-\phi+\frac{4\pi}{3},
\phi+\frac{4\pi}{3}\right].
\end{equation} 
The integrand in Eq.~\eref{prtheta} can be expanded as a sum of 
$\cos k\theta$ and $\sin k\theta$. Terms of the form $\sin k\theta$
yield zero upon integration; terms of the form $\cos k\theta$ yield 
\begin{equation}
\int_{\calc_{\phi}}d\theta \cos k\theta=\int_{-\phi}^{\phi}d\theta 
\left(1+\cos\frac{2k\pi}{3}+\cos\frac{4k\pi}{3}\right)\cos k\theta,
\end{equation}
which is zero unless $3|k$. Nonzero contributions therefore
necessarily come from the $\cos\left(\theta+\frac{4\pi}{3}\right)
\cos^2\left(\theta+\frac{2\pi}{3}\right)\sin(3\theta)$ term in \eref{prtheta}.
Keeping only terms of the form $\cos 3kt$ in the expansion of this
term we get
\begin{equation}
\label{prrphi}
P(R)=70\sqrt{3}r^6\left(2\pi-6\phi+\sin(6\phi)\right).
\end{equation}
Replacing $\phi$ and $R$ by their value we finally obtain
\begin{equation}
\label{pr33}
\begin{array}{c|l}
P(R)=70\sqrt{3}&2\pi\left(R-\frac{1}{3}\right)^3, \hspace{3cm} 
\frac{1}{3}\leq R\leq \frac{1}{2}\\ 
&6\left(R-\frac{1}{3}\right)^3\left(\frac{\pi}{3}-\arccos\frac{1}{\sqrt{6R-2}}\right)\\
&\ \ \ \ \ \ +(R-1)(R-\frac{5}{9})\sqrt{6R-3},\ \ \  \frac{1}{2}\leq R\leq 1.
\end{array}
\end{equation}
It is interesting to check whether this formula allows to recover the moments 
derived in \cite{Gir07}. The value of $\langle R^n \rangle$ is given 
(see  \cite{Gir07}) by 
\begin{eqnarray}
\label{rnferme}
\langle R^n \rangle=\frac{p!(p q-1)!}{(p q+2n-1)!}
\sum_{n_1+n_2+\cdots+n_p=n}\frac{n!}{n_1!n_2!\ldots n_p!}\nonumber\\
\hspace{2cm}\times
\prod_{i=1}^{p}\frac{(q+2n_i-i)!}{(q-i)!i!}
\prod_{1\leq i<j\leq p}(2n_i-i-2n_j+j)
\end{eqnarray}
(correcting by a factor $p!$ the Eq. (10) of \cite{Gir07}, where this 
term had been erroneously forgotten). 
The first moments of $P(R)$ as calculated from Eq.~\eref{pr33} yield values that 
are precisely equal to those obtained from Eq.~\eref{rnferme}.

\section{Distribution of the purity for a 
$3 \times q$ bipartite random state, $q\geq 3$.\label{pder3q}}
In the case of a random vector belonging to a Hilbert space of
dimensions $p\times q$ with $p=3$ and $q\geq 3$, 
the term $V(x_1, x_2, x_3)x_2 x_3^2$ in \eref{defpr3}
is multiplied by a factor $(x_1x_2x_3)^d$, where $d=q-p$. The changes of
variables of Section \ref{pder33} yield (still setting
$r=\sqrt{R-1/3}$) an integrand that 
can be expanded as
\begin{eqnarray}
\label{develop}
&-\frac{r^3}{\sqrt{2}}\sum_{k=0}^{d}\bino{d}{k}
\left(\frac{1}{27}-\frac{r^2}{6}\right)^{d-k}
\left(-\frac{r^3}{3\sqrt{6}}\right)^{k}
\left(\frac{1}{3}-r\sqrt{\frac{2}{3}}\cos\left(\theta+\frac{4\pi}{3}\right)
\right)\nonumber\\
&\hspace{2.5cm}\times
\left(\frac{1}{3}-r\sqrt{\frac{2}{3}}\cos\left(\theta+\frac{2\pi}{3}
\right)\right)^2\sin 3\theta\cos^k 3\theta.
\end{eqnarray}
Again in \eref{develop} only terms of the form
$\cos(3j\theta)$ contribute to the final result, and nonzero contributions therefore
necessarily come from the 
\begin{equation}
f_k(\theta)=\cos\left(\theta+\frac{4\pi}{3}\right)
\cos^2\left(\theta+\frac{2\pi}{3}\right)\sin(3\theta)\cos^k 3\theta
\end{equation}
term. For a given $k$, the expansion of $\cos^k 3\theta$ yields
\begin{equation}
\cos^k 3\theta=\left\{\begin{array}{ll}
\frac{1}{2^{k-1}}\sum_{j=0}^{k/2}\left(1-\frac{1}{2}\delta_j\right)
\bino{k}{k/2-j}\cos6j\theta&k\textrm{ even}\\
\frac{1}{2^{k-1}}\sum_{j=0}^{(k-1)/2}
\bino{k}{(k-1)/2-j}\cos(6j+3)\theta&k\textrm{ odd,}
\end{array}\right.
\end{equation}
where $\delta_j$ is the Kronecker delta. 
For a fixed integer $t$ the only terms contributing to $P(R)$ in the expansion of  
$\cos\left(\theta+\frac{4\pi}{3}\right)\cos^2\left(\theta+\frac{2\pi}{3}\right)
\sin(3\theta)\cos 3t\theta$ sum up to
$\sqrt{3}\left(\cos(3t-6)\theta-2\cos 3t\theta+\cos(3t+6)\theta\right)/32$.
The integral of $f_k(\theta)$ over $\calc_{\phi}$ thus yields
\begin{eqnarray}
\label{kpair}
\frac{3\sqrt{3}}{2^{k+3}}\sum_{j=0}^{k/2}\left(1-\frac{1}{2}\delta_j\right)
\bino{k}{k/2-j}\chi_j(\phi),\hspace{1cm}
&k\textrm{ even}\\
\label{kimpair}
\frac{3\sqrt{3}}{2^{k+3}}\sum_{j=0}^{(k-1)/2}
\bino{k}{(k-1)/2-j}\chi_{j+1/2}(\phi),
&k\textrm{ odd,}
\end{eqnarray}
where we have defined, for $j\geq 0$, the function 
\begin{equation}
\chi_j(\phi)=\frac{\sin(6j-6)\phi}{6j-6}-2\frac{\sin6j\phi}{6j}
+\frac{\sin(6j+6)\phi}{6j+6}
\end{equation}
(for $j=0$ and $j=1$ it is understated that the limit $j\to 0$ or $j\to 1$
is taken). Summing all contributions together 
we finally obtain the exact expression
\begin{eqnarray}
\label{prfinal3q}
P(R)=\frac{(3q-1)!}{16\sqrt{3}\prod_{j=1}^{3}(q-j)!}
\sum_{k=0}^{d}\bino{d}{k}\left(\frac{-1}{6\sqrt{6}}\right)^k
\left(\frac{5-9R}{54}\right)^{d-k}
\left(R-\frac{1}{3}\right)^{\frac{3}{2}(k+2)}\nonumber\\
\times\sum_{j=0}^{\lfloor k/2\rfloor}
\left(1-\frac{1}{2}\delta_j\delta_{\bar{k}}\right)
\bino{k}{\lfloor \frac{k}{2}\rfloor-j}\left(\chi_{j+\bar{k}/2}(\phi)
-\chi_{j+\bar{k}/2}(\frac{\pi}{3})\right),
\end{eqnarray}
with again $\phi=0$ for  $1/3\leq R\leq 1/2$ and $\arccos(1/\sqrt{6R-2})$ for
 $1/2\leq R\leq 1$.
Here $\lfloor x\rfloor$ is the integer part of 
$x$,  $\bar{k}=k\bmod 2$ and $\delta$ is the Kronecker delta symbol. 
Note that the term $\chi_{j+\bar{k}}(\phi)$ can be expressed
as a function of $R$ in terms of Chebychev polynomials of the second kind $U_n$, using
the relation $\sin n \theta=U_{n-1}(\cos\theta)\sin\theta$.
Again, on can check that the expression \eref{prfinal3q} allows to 
recover the moments \eref{rnferme}.

\section{Distribution of the purity for a 
$4 \times 4$ bipartite random state. \label{pder44}}
The treatment of the $p=4$ case is quite similar to the previous one
but calculations (and results) quickly get very heavy. Here we derive an explicit 
expression for the case $p=q=4$. The first steps of Section
\ref{pder33} are easily generalized (they can in fact be generalized in an obvious
way to arbitrary $p$, $q$). The domain of integration in \eref{defpr} is the
intersection of a plane and a hypersphere in $\mathbb{R}^4$, restricted to
$[0,1]^4$.
First we make a change of variables 
${\bf x}=O{\bf X}$, where $O$ is the orthogonal matrix
\begin{equation}
\label{matrixO4}
\left(
\begin{array}{cccc}
-\frac{3}{\sqrt{12}}&0&0&\frac{1}{2}\\
\frac{1}{\sqrt{12}}&-\frac{2}{\sqrt{6}}&0&\frac{1}{2}\\
\frac{1}{\sqrt{6}}&\frac{1}{\sqrt{6}}&-\frac{1}{\sqrt{2}}&\frac{1}{2}\\
\frac{1}{\sqrt{6}}&\frac{1}{\sqrt{6}}&\frac{1}{\sqrt{2}}&\frac{1}{2}
\end{array}\right),
\end{equation}
so that the plane has equation $X_4=1/2$ and the sphere
$X_1^2+X_2^2+X_3^2+X_4^2=R$. After integration over $X_4$, we have to integrate 
\begin{equation}
\label{aintegrer}
f(X_1, X_2, X_3)=x_2 x_3^2 x_4^3 V(x_1, x_2, x_3, x_4)
\end{equation}
over the portion of the sphere of radius $r=\sqrt{R-1/4}$ that is comprised
inside the thetrahedron of vertices $V_1=(-\sqrt{3}/2,0,0)$, 
$V_2=(1/\sqrt{12},-2/\sqrt{6},0)$, $V_3=(1/\sqrt{12},1/\sqrt{6}, -1/\sqrt{2})$
and $V_4=(1/\sqrt{12},1/\sqrt{6}, 1/\sqrt{2})$, centered on $(0,0,0)$.
In the case $1/4<R<1/3$, the sphere is entirely inside the tetrahedron:
after making a change of variables 
to spherical coordinates $(X_1=\rho\cos\theta_1,
X_2=\rho\sin\theta_1\cos\theta_2,X_3=\rho\sin\theta_1\sin\theta_2)$, the
integration is trivial. Taking into account the factors 
$1/2$ and $1/2r$ coming from the integration 
of $\delta(2X_4-1)$ and $\delta(R-1/4-\rho^2)$ the integration of 
\eref{aintegrer} yields 
\begin{equation}
\label{pr44cas1}
\int d{\bf X} f({\bf X})=4\pi r^{13}/45045,
\end{equation}
with ${\bf X}=(X_1, X_2, X_3)$.
When  $1/3<R<1/2$, the integrand can first be simplified by use of the symmetries of
the domain of integration, which consists of the sphere minus 
the four spherical caps emerging from the tetrahedron.
We note $C_i$ , 
$1\leq i\leq 4$ the cap opposite to vertex $V_i$. Let $R_i(\varphi)$ be the
rotation of axis $O X_i$, $i=1,2,3$, and angle $\varphi$. 
The caps $C_2, C_3, C_4$ can be obtained from $C_1$ by rotations: 
$C_2$ is the image of $C_1$ by 
$S=R_1(-\pi/3)R_3(\arccos(-1/3))$, and $C_3$ and $C_4$ are respectively the
images of $C_2$ by $T=R_1(2\pi/3)$ and  $T^2=R_1(4\pi/3)$. Therefore
\begin{equation}
\label{int4C}
\int_{\bigcup_{i=1}^{4}C_i}\hspace{-0.5cm}d{\bf X} f({\bf X})=
\int_{C_1}d{\bf X}\left(f({\bf X})+f(S{\bf X})+f(TS{\bf X})+f(T^2S{\bf X})\right).
\end{equation}
In spherical 
coordinates, the top cap (cap $C_1$)
has equation $\rho=r, 0\leq \theta_1\leq
\arccos\frac{1}{2r\sqrt{3}}, 0\leq \theta_2\leq 2\pi$ (we recall that  $r=\sqrt{R-1/4}$).
Performing the integrals over $\theta_2$, and then $\theta_1$, in \eref{int4C} 
yields
\begin{eqnarray}
\label{pr44cas2}
\int d{\bf X} f({\bf X})= &\frac{967 \pi }{804925734912 \sqrt{3}} - 
\frac{571 \pi}{9459597312 \sqrt{3}}r^2 + 
\frac{505 \pi}{429981696\sqrt{3}}r^4\nonumber\\
&\hspace{-4.5cm}- \frac{229 \pi}{20901888 \sqrt{3}}r^6
+ \frac{\pi} {20480 \sqrt{3}}r^8- \frac{11 \pi}{124416 \sqrt{3}}r^{10} 
- \frac{\pi}{20736\sqrt{3}}r^{12} +\frac{8\pi}{45045}r^{13}.
\end{eqnarray}
Finally, when $1/2<R<1$, the domain of integration can be further reduced to 
$\theta_2\in[0, \pi/3]$ by applying $T$ and $T^2$ and using the
symmetry $\theta_2\to-\theta_2$. 
The domain of integration is $\theta_2\in[0, \pi/3]$ and
$\theta_1\in [\varphi, \pi[$ where $\varphi$ verifies
\begin{equation}
\label{limitdomain}
\cos\theta_2=\frac{r\cos\varphi+\sqrt{3}/2}{2\sqrt{2}r\sin\varphi},
\end{equation}
and the integrand is a sum of terms of the form 
$\cos^{2k}\theta_1\sin\theta_1\cos^{2k'}\theta_2$ and
$\cos^{2k+1}\theta_1\cos^{2k'+1}\theta_2$.
The calculation gets tedious and the steps leading to $P(R)$ in this case 
are given in the Appendix.

Taking into account the normalization factor $\cala p!=15!/(3!2!)^2$
we finally get $P(R)=0$ for $R\notin[1/4,1]$ and
\begin{eqnarray}
\label{pr44}
P(R)&=&\frac{1575\pi}{16}\left(4R-1\right)^{\frac{13}{2}},
\hspace{3cm}  R\in\left[\frac{1}{4},\frac{1}{3}\right]\\
P(R)&=&\frac{\pi}{3}Q_2(R)-
\frac{1575\pi}{16}\left(4R-1\right)^{\frac{13}{2}},
\hspace{1.3cm}  R\in\left[\frac{1}{3},\frac{1}{2}\right]\nonumber\\
P(R)&=&\sqrt{6R-3}\ Q_1(R)+\left(\frac{\pi}{3}
-\arccos\left(\frac{1}{\sqrt{6R-2}}\right)\right)Q_2(R)\nonumber\\
&-&\frac{4725}{16}(4R-1)^{\frac{13}{2}}
\left(\frac{\pi}{3}-\arccos\left(\frac{R}{3R-1}\right)\right),\nonumber
R\in\left[\frac{1}{2},1\right]\nonumber
\end{eqnarray}
where we have introduced the following two polynomials
$Q_1(x)=\frac{175}{1944\sqrt{3}}(1-x)(1657 
+ 277731 x - 2190321 x^2 + 6208416 x^3 - 7386066 x^4 +  2913408 x^5)$
and $Q_2(x)=\frac{175}{1944\sqrt{3}}(-159241 + 2178306 x - 11709126 x^2 + 30254796 x^3 
- 34540506 x^4 + 4864860 x^5 + 14594580 x^6)$. 
Again it can be checked analytically that the moments obtained from the above 
formula for $P(R)$ agree with those obtained from Eq.~\eref{rnferme}.

\section{Probability density distribution of the purity for a 
$4 \times q$ bipartite random pure state, $q\geq 4$. \label{pder4q}}
Similar calculations can be done in the same way for $q>4$. 
However as the number of terms in the integrand gets large, it is more 
efficient to proceed in the following
way. The only difference between the $q=4$ and the $q>4$ cases is that the integrand
is multiplied by $(x_1 x_2 x_3 x_4)^d$, where $d=q-p$.
It is then easy to see from Section \ref{pder44}
that the integrals appearing in the calculation for $1/4<R<1/3$ 
and $1/3<R<1/2$ still yield polynomials in $r$. For $1/2<R<1$, 
one can show that after reducing the domain of integration by symmetries as in the 
previous section, the integrand is again a sum of terms of the form 
$\cos^{2k}\theta_1\sin\theta_1\cos^{2k'}\theta_2$ and
$\cos^{2k+1}\theta_1\cos^{2k'+1}\theta_2$. 
The calculation therefore yields an expression similar 
to Eq.~\eref{pr44cas3}, but with polynomials in $r$ of higher degree.
Therefore one can look for a $P(R)$ (expressed in terms of $r=\sqrt{R-1/4}$) 
of the form $A_5(r)$ for $1/4\leq R\leq 1/3$, $A_6(r)$ for $1/3\leq R\leq 1/2$,
and of the form of Eq.~\eref{pr44cas3} for $1/2\leq R\leq 1$,
where $A_i$ all are polynomials in $r$. The maximal order of the $A_i$ 
corresponds to the order of the integrand 
$(x_1x_2x_3x_4)^d x_2 x_3^2 x_4^3 V(x_1,x_2,x_3,x_4)$, that 
is (taking into account the $r^2$ from the Jacobian 
and the $1/r$ of the integration of one of the delta functions) 
$13+4d$. The coefficients of these
polynomials are unknown variables.
The knowledge of all moments \eref{rnferme} allows
us to write down as many linear equations as there are unknown variables, that 
is $6(14+4d)$. 
The problem now reduces to solving a linear system $Mx=b$. The vector
$b=\{\langle R^0\rangle, \langle R^1\rangle, \langle R^2\rangle,
\ldots\}$ is obtained from \eref{rnferme}, 
and $M$ is a matrix of size  $6(14+4d)$ 
whose coefficients are given by terms of the form
\begin{equation}
\int_{\sqrt{1/k-1/4}}^{\sqrt{1/(k-1)-1/4}}
r^{\nu}g(r)\ dr,\ \ \ \ \ \ \ \ k=2,3,4,
\end{equation}
where $g(r)$ is one of the functions $1$,
$\arctan\sqrt{\frac{8r^2}{4r^2-1}}$,
$\arctan\sqrt{\frac{2}{12r^2-3}}$ or $\sqrt{4r^2-1}$.
These integrals can be calculated analytically. The solution
$x=M^{-1}b$ of the system yields the coefficients of the polynomials
$A_i$ and thus an analytic formula for $P(R)$.

A similar method would yield expressions for higher Hilbert space dimensions.
However, given what Eq.~\eref{pr44} looks like, it is not clear whether
formulae for higher dimensions could be cast into a tractable form.\\

The author thanks CalMiP in Toulouse and Idris in Orsay for access to their 
supercomputers, and Phuong Mai Dinh for helpful comments.
This work was supported by the Agence Nationale de 
la Recherche (ANR project INFOSYSQQ) and the European program 
EuroSQIP.

\section*{Appendix}
The derivation of $P(R)$ for $p=4, q=4$ and $\frac{1}{2}\leq R\leq 1$ starts from the integrand
obtained after having reduced the domain of integration by symmetries, 
and integrated over $\rho$ and $\theta_1\in[\varphi, \pi[$.
From Eq.~\eref{limitdomain} and using the fact that
$\theta_2\in[0,\pi/3[$ we obtain an expression for
$\sin\varphi$ and $\cos\varphi$ which allows to express
the integrand as a function of $\theta_2$ only.
Expanding all trigonometric terms and changing
variables from $\cos\theta_2$ to $x$, we are left with
a sum of terms of the form
\begin{equation}
\label{type1}
\frac{x^{2a}\sqrt{1-x^2}}{(1+8x^2)^{13}}
\end{equation}
and
\begin{equation}
\label{type2}
\frac{x^{2a+1}\sqrt{1-x^2}\sqrt{4r^2(1+8x^2)-3}}{(1+8x^2)^{13}},
\end{equation}
where $a$ is some integer and $x$ has to be 
integrated between $1/2$ and $1$. 
The integrals corresponding to \eref{type1} give constants independent of $r$.
The integrals corresponding to \eref{type2} can be evaluated by the change of
variables $t=\sqrt{(x^2-c)/(1-x^2)}$ with $c=(3-4r^2)/(32r^2)$. 
Upon integration, we obtain terms of the form
\begin{equation}
\label{pr44cas3}
\hspace{-2cm}A_1(r)+A_2(r)\arctan\sqrt{\frac{8r^2}{4r^2-1}}+A_3(r)\arctan\sqrt{\frac{2}{12r^2-3}}+A_4(r)\sqrt{4r^2-1},
\end{equation}
where $A_i$ are polynomials in $r$. Replacing $r$ by its value
$\sqrt{R-\frac{1}{4}}$ and noting that
\begin{eqnarray}
\arccos\sqrt{\frac{2}{12r^2-1}}+\arctan\sqrt{\frac{2}{12r^2-3}}&=&\frac{\pi}{2}\nonumber\\
\frac{1}{2}\arccos\frac{4r^2+1}{12r^2-1}+\arctan\sqrt{\frac{8r^2}{4r^2-1}}&=&\frac{\pi}{2},
\end{eqnarray}
we get the final result \eref{pr44}.\\

\end{document}